\documentclass{ptephy_v1}
\usepackage{subcaption}

\begin{document}

\title{Hidden Nambu mechanics II: 
{\LARGE Quantum/semiclassical dynamics}}


\author{\fname{Atsushi} \surname{Horikoshi}}
\affil{Department of Natural Sciences, Tokyo City University,
Tokyo 158-8557, Japan \email{horikosi@tcu.ac.jp}}


\begin{abstract}%
Nambu mechanics is a generalized Hamiltonian dynamics characterized by
an extended phase space and multiple Hamiltonians.
In a previous paper [Prog. Theor. Exp. Phys. {\bf 2013}, 073A01 (2013)]
we revealed that the Nambu mechanical structure is hidden in Hamiltonian dynamics,
that is, the classical time evolution of variables including redundant degrees of freedom
can be formulated as Nambu mechanics.
In the present paper
we show that the Nambu mechanical structure is also hidden in some quantum or semiclassical dynamics,
that is, in some cases the quantum or semiclassical time evolution of expectation values of
quantum mechanical operators, including composite operators,
can be formulated as Nambu mechanics.
We present a procedure to find hidden Nambu structures in quantum/semiclassical systems 
of one degree of freedom, and give two examples: 
the exact quantum dynamics of a harmonic oscillator, and semiclassical wave packet dynamics.
Our formalism can be extended to many-degrees-of-freedom systems;
however, there is a serious difficulty in this case due to interactions between degrees of freedom.
To illustrate our formalism we present two sets of numerical results on semiclassical dynamics:
from a one-dimensional metastable potential model and 
a simplified Henon--Heiles model of two interacting oscillators.
\end{abstract}

\subjectindex{A30, A60}

\maketitle
\section{Introduction}
\label{Introduction}

In 1973, Nambu proposed a generalization of the classical Hamiltonian dynamics \cite{Nambu}
that is nowadays referred to as the Nambu mechanics.
In his formulation, the phase space spanned by the canonical doublet $(q, p)$
is extended to that spanned by $N(\ge 3)$ variables $(x_1, x_2, ..., x_{N})$, the Nambu $N$-plet,
and the Hamilton equations of motion are generalized to the Nambu equations.
In order for the Liouville theorem to hold in the $N$-dimensional extended phase space,
the Nambu equations are defined by $N-1$ Nambu Hamiltonians and the Nambu bracket, 
an $N$-ary generalization of the Poisson bracket. 
The structure of Nambu mechanics has impressed many authors,
who have reported studies on its fundamental properties and possible applications,
including quantization of the Nambu bracket 
\cite{Takhtajan,DitoFlatoSternheimerTakhtajan,AwataLiMinicYoneya,CurtrightZachos,AxenidesFloratosNicolis,
deAzcarragaIzquierdo,MongkolsakulvongChaikhanFrank,BlenderLucarini,SaitouBambaSugamoto,HoMatsuo,Yoneya}.
However, the applications to date have been limited to particular systems,
because Nambu systems generally require multiple conserved quantities as Hamiltonians and 
the Nambu bracket exhibits serious difficulties in systems with many degrees of freedom 
or quantization \cite{Nambu,Takhtajan,HoMatsuo}.
\par
In 2013 we proposed a new approach to Nambu mechanics \cite{HorikoshiKawamura}.
We revealed that the Nambu mechanical structure is hidden in a Hamiltonian system 
which has redundant degrees of freedom.
For example, in a Hamiltonian system with a Hamiltonian $H(q,p)$,
if we take three variables as $(x_1, x_2, x_3)=(q,p,q^2)$,
their classical time evolution can be given by $N$ = 3 Nambu equations
with two Hamiltonians $F(x_1, x_2, x_3)$ and $G(x_1, x_2, x_3)$.
Here $x_3=q^2$ is a redundant degree of freedom in the original Hamiltonian system,
and the Nambu Hamiltonians are given by
the original Hamiltonian $F(x_1, x_2, x_3)=H(q,p)$ and 
the constraint $G(x_1, x_2, x_3)=x_3-x_1^2=0$, which is induced due to the consistency between the three variables. 
We derived the consistency condition to determine the induced constraints.
\par
In the present paper we show that the Nambu mechanical structure is also hidden 
in some quantum or semiclassical systems.
The key idea is as follows.
In our previous work, the Nambu multiplet is given as a function of classical variables $(q,p)$,
and therefore the induced constraints are always trivial, i.e. set to zero \cite{HorikoshiKawamura}. 
However, if we take the Nambu multiplet as a set of expectation values of
quantum mechanical operators including composite operators $(\hat{q}^{2},\hat{p}^{2},...)$, 
the constraints become nontrivial because of quantum fluctuation. 
Furthermore, if these constraints are constants of motion,
the time evolution of the Nambu multiplet could be given by the Nambu equations.
For example, consider a classical system with a Hamiltonian $H(q,p)$ and 
a corresponding quantum system with the Hamiltonian $\hat{H}=H(\hat{q},\hat{p})$.
If we take three variables as $(x_1, x_2, x_3)=(q,p,q^2)$,
the trivial constraint $G=x_3-x_1^2=0$ is induced.
Then, if we replace these variables with  
$(x_1, x_2, x_3)=(\langle\hat{q}\rangle,\langle\hat{p}\rangle, \langle\hat{q}^{2}\rangle)$,
the same function $G=x_3-x_1^2$ has a nonzero value in general because of quantum fluctuation.
Furthermore, in the case of frozen Gaussian wave packet dynamics \cite{Heller},
which is the dynamics of a Gaussian wave packet with a fixed width $\sigma$,
the function $G=x_3-x_1^2=\sigma^2$ is constant in time 
and therefore the quantum or semiclassical time evolution of the Nambu triplet
can be given by the $N$=3 Nambu equations with Nambu Hamiltonians $F$ and $G$.
Here, $F$ is equal to or approximately equal to the expectation value of the Hamiltonian operator
$F=\langle\hat{H}\rangle$ or $F\simeq\langle\hat{H}\rangle$.
We present a general procedure to find the Nambu mechanical structure in quantum or semiclassical systems
of one degree of freedom with some specific examples.  
It should be noted that our formulation is {\it not} a quantization of the Nambu bracket.
We just propose a prescription to describe ordinary quantum or semiclassical dynamics 
in a classical Nambu mechanical manner.  
\par
The Nambu mechanical structure is hidden not only in one-degree-of-freedom systems.
It is straightforward to extend our formalism to many-degrees-of-freedom systems 
by extending the definition of the Nambu bracket.
However, the resulting hidden Nambu mechanics becomes pathological
because in many-degrees-of-freedom systems the Nambu bracket does not satisfy the fundamental identity, 
which is an important property of the Nambu bracket and
corresponds to the Jacobi identity in Hamiltonian dynamics \cite{Takhtajan,HoMatsuo}.
Without the Jacobi identity 
the canonical transformation of the canonical doublet cannot be properly defined,
and the dynamics becomes anomalous \cite{SatoYoshida}.
Similarly, without the fundamental identity
we cannot define the canonical transformation of Nambu multiplets 
including their consistent time evolution. 
The hidden Nambu mechanics in many-degrees-of-freedom systems is
an example of dynamics without the canonical structure.
\par
The outline of this article is as follows.
In Sect. 2 
we briefly review our previous work on
the hidden Nambu mechanics in classical Hamiltonian systems \cite{HorikoshiKawamura}.
As preparation for the next section, we give a detailed description of some examples. 
In Sect. 3 we present a procedure to find the Nambu mechanical structure hidden 
in some quantum or semiclassical dynamics.
Two examples are given:
the exact quantum dynamics of a harmonic oscillator and 
the semiclassical nonlinear dynamics of a frozen Gaussian wave packet.
In Sect. 4 we give an extension of our formalism to many degrees of freedom.
In Sect. 5 we present two numerical results to illustrate our formalism:
the semiclassical tunneling dynamics in a one-dimensional metastable system and
the semiclassical energy exchange dynamics between two coupled oscillators in a simplified Henon--Heiles model.
In the last section we give our conclusions and discuss the direction of future work.

\section{Hidden Nambu mechanics}
\label{Hidden Nambu mechanics}
We begin with a brief review of Hamiltonian dynamics, Nambu mechanics \cite{Nambu}, and
hidden Nambu mechanics \cite{HorikoshiKawamura}.
We describe two examples in detail to prepare for the next section. 
In this and the next section we consider only one-degree-of-freedom systems.   
\subsection{Hamiltonian dynamics}
\label{Hamiltonian dynamics}
Hamiltonian dynamics is the classical dynamics of the canonical doublet $(q(t),p(t))$, 
which is given by a Hamiltonian $H=H(q, p)$ and the Poisson bracket 
defined by the two-dimensional Jacobian,
\begin{eqnarray}
\{A, B\}_{\mbox{\tiny{PB}}} \equiv  
\frac{\partial (A,B)}{\partial (q,p)}=
\frac{\partial A}{\partial q}\frac{\partial B}{\partial p}
 - \frac{\partial A}{\partial p}\frac{\partial B}{\partial q},
\label{PB}
\end{eqnarray}
where $A=A(q,p)$ and $B=B(q,p)$ are any functions of the canonical doublet. 
The Poisson bracket should satisfy the Jacobi identity,
\begin{eqnarray}
\{\{A_{1}, A_{2}\}_{\mbox{\tiny{PB}}},B\}_{\mbox{\tiny{PB}}}= 
\{\{A_{1}, B\}_{\mbox{\tiny{PB}}},A_{2}\}_{\mbox{\tiny{PB}}}+
\{A_{1},\{A_{2}, B\}_{\mbox{\tiny{PB}}}\}_{\mbox{\tiny{PB}}},
\label{JI}
\end{eqnarray}
where $A_{1}=A_{1}(q,p)$, $A_{2}=A_{2}(q,p)$, and $B=B(q,p)$ are any functions.
In terms of the Poisson bracket,
the Hamilton equation of motion for any function $f=f(p,q)$ can be written as
\begin{eqnarray}
\frac{d f}{dt} = \{f, H\}_{\mbox{\tiny{PB}}}.
\label{H-eq}
\end{eqnarray}
The time evolution according to this equation preserves the phase space volume
because of the divergenceless property,
\begin{eqnarray}
\frac{\partial }{\partial q}\frac{dq}{dt} + \frac{\partial }{\partial p}\frac{dp}{dt} =
\frac{\partial }{\partial q}\{q,H \}_{\mbox{\tiny{PB}}}+\frac{\partial }{\partial p}\{p,H \}_{\mbox{\tiny{PB}}}=0.
\label{divH}
\end{eqnarray}
This is the Liouville theorem in Hamiltonian dynamics.

\subsection{Nambu mechanics}
\label{Nambu mechanics}
Nambu mechanics is a generalized Hamiltonian dynamics
of $N(\ge 3)$ variables $(x_1, x_2,..., x_{N})$ \cite{Nambu}. 
In Nambu mechanics 
the canonical doublet is generalized to the Nambu $N$-plet,
and the Poisson bracket, Eq. (\ref{PB}), is generalized to the Nambu bracket 
defined by means of the $N$-dimensional Jacobian,
\begin{eqnarray}
\{{A}_1, {A}_2, ..., {A}_N\}_{\mbox{\tiny{NB}}} 
&\equiv& 
\frac{\partial ({A}_1, {A}_2, ..., {A}_N)}{\partial (x_1, x_2, ..., x_N)}
\nonumber \\
&=& \sum_{i_1, i_2, ..., i_N=1}^{N} \varepsilon_{i_1 i_2 \cdots i_N}
\frac{\partial {A}_1}{\partial x_{i_1}}\frac{\partial {A}_2}{\partial x_{i_2}}
 ... \frac{\partial {A}_{N}}{\partial x_{i_N}},
\label{NB}
\end{eqnarray}
where ${A}_{a}={A}_{a}(x_1, x_2, ..., x_N)$ $(a=1, ..., N)$
are any functions of the Nambu multiplet and 
$\varepsilon_{i_1 i_2 \cdots i_{N}}$ is the $N$-dimensional Levi--Civita symbol, 
the antisymmetric tensor with $\varepsilon_{12 \cdots N} =1$.
The Nambu bracket should satisfy the following fundamental identity \cite{Takhtajan},
an $N$-ary generalization of the Jacobi identity in Eq. (\ref{JI}): 
\begin{eqnarray}
\{\{A_{1},..., A_{N}\}_{\mbox{\tiny{NB}}}, B_{1},..., B_{N-1}\}_{\mbox{\tiny{NB}}}= 
\sum_{a=1}^{N}
\{A_{1},...,\{A_{a}, B_{1},..., B_{N-1}\}_{\mbox{\tiny{NB}}},..., A_{N}\}_{\mbox{\tiny{NB}}},
\label{FI}
\end{eqnarray}
where $B_{b}=B_{b}(x_1, x_2, ..., x_N)$ $(b=1, ..., N-1)$ are any functions.
In terms of the Nambu bracket, 
the Nambu equation for any function ${f}={f}(x_1, x_2, ..., x_N)$
can be written as
\begin{eqnarray}
\frac{d {f}}{dt} = \{{f}, {H}_1, ..., {H}_{N-1}\}_{\mbox{\tiny{NB}}},
\label{N-eq}
\end{eqnarray}
where ${H}_b={H}_b(x_1, x_2, ..., x_N)$ $(b=1, ..., N-1)$ are Nambu Hamiltonians.
The time evolution according to this equation preserves the $N$-dimensional phase space volume
because of the divergenceless property,
\begin{eqnarray}
\sum_{i=1}^{N}\frac{\partial }{\partial x_{i}}\frac{dx_{i}}{dt} =
\sum_{i=1}^{N}\frac{\partial }{\partial x_{i}}\{x_{i}, {H}_1, ..., {H}_{N-1}\}_{\mbox{\tiny{NB}}}=0.
\label{divN}
\end{eqnarray}
Therefore the Liouville theorem also holds in Nambu mechanics.

\subsection{Hidden Nambu mechanics}
\label{Hidden Nambu mechanics-c}

Consider a Hamiltonian system of a canonical doublet $(q, p)$ with a Hamiltonian $H=H(q, p)$.
The key idea of hidden Nambu mechanics is to describe this system
by means of $N(\ge 3)$ variables $x_i=x_i(q, p)$ $(i=1, ..., N)$.
We assume that at least $N-1$ of 
$\{x_i, x_j\}_{\mbox{\tiny{PB}}}$ do not vanish,
so that the time evolution of any function 
$\tilde{f}(x_1, ..., x_N) = f(q, p)$
can be written via Hamilton equation of motion in Eq. (\ref{H-eq}), 
\begin{eqnarray}
\frac{d \tilde{f}}{dt} = \frac{d f}{dt} = \frac{\partial(f, H)}{\partial(q, p)} 
= \frac{1}{2} \sum_{i, j=1}^{N} 
\frac{\partial (\tilde{f}, F)}{\partial (x_{i}, x_{j})}
\{x_{i}, x_{j}\}_{\mbox{\tiny{PB}}},
\label{H-eq2}
\end{eqnarray}
where $F(x_1, ..., x_N) = H(q, p)$.
We introduce the functions $G_{c}=G_{c}(x_1, ..., x_N)$ $(c=1, ..., N-2)$
which satisfy the consistency conditions
\begin{eqnarray}
\frac{1}{(N-2)!} \sum_{k_1, ..., k_{N-2}=1}^{N}
\varepsilon_{i j k_1 \cdots k_{N-2}}
\frac{\partial (G_1, ..., G_{N-2})}
{\partial (x_{k_1}, ..., x_{k_{N-2}})}
= \{x_{i}, x_{j}\}_{\mbox{\tiny{PB}}}.
\label{condition}
\end{eqnarray}
Then, Eq. (\ref{H-eq2}) can be rewritten as the Nambu equation in the form of Eq. (\ref{N-eq}),
\begin{eqnarray}
\frac{d \tilde{f}}{dt} = 
\{\tilde{f}, F, G_1, ..., G_{N-2}\}_{\mbox{\tiny{NB}}},
\label{HN-eq}
\end{eqnarray}
where we have used the following formula concerning Jacobians:
\begin{eqnarray}
\frac{\partial (A_1, A_2, ..., A_N)}
{\partial (x_1, x_2, ..., x_N)}
= \frac{1}{2(N-2)!} \sum_{i_1, ..., i_{N}=1}^{N}
\varepsilon_{i_1 \cdots i_{N}}
\frac{\partial (A_1, A_2)}{\partial (x_{i_1}, x_{i_2})}
\frac{\partial (A_3, ..., A_{N})}
{\partial (x_{i_3}, ..., x_{i_{N}})}.
\label{J(gN)}
\end{eqnarray}
The functions $G_c$ are constants in motion and 
can be set to zero by redefining $G_c$.
This is a natural choice because the functions $G_c$
work as constraints for the Nambu multiplet $(x_1, x_2, ..., x_N)$.
We refer to $G_c=0$ as {\it induced constraints}, because they are induced 
by enlarging the phase space from $(q, p)$ to $(x_1,x_2,..., x_{N})$.

\subsection{Examples}
\label{Examples}

Here we present detailed descriptions of two simple examples to show how induced constraints
are obtained for given multiplets.
We adopt the same choice of $N$-plets in the next section.
Finally we comment on the functional forms of the Nambu Hamiltonians.
\\

{\bf (a) $N=3$: classical harmonic oscillator}\\
Consider three composite variables of the canonical doublet,
\begin{eqnarray}
x_{1}=q^2,~~~ 
x_{2}=p^2,~~~ 
x_{3}=qp,
\label{Ex1}
\end{eqnarray}
which satisfy the following relations:
\begin{eqnarray}
\{x_{1}, x_{2}\}_{\mbox{\tiny{PB}}} = 4x_{3},~~~
\{x_{2}, x_{3}\}_{\mbox{\tiny{PB}}} = -2x_{2},~~~ 
\{x_{3}, x_{1}\}_{\mbox{\tiny{PB}}} = -2x_{1}.
\label{Ex1-PB}
\end{eqnarray}
Then, the conditions in Eq. (\ref{condition}) become
\begin{eqnarray}
\frac{\partial G}{\partial x_{1}} = -2x_{2},~~~ 
\frac{\partial G}{\partial x_{2}} = -2x_{1},~~~ 
\frac{\partial G}{\partial x_{3}} = 4x_{3},
\label{Ex1-dG}
\end{eqnarray} 
and $G$ is solved as $G= 2x_{3}^{2}-2x_{1}x_{2}+C$ with a constant $C$.
Redefining $G$ to eliminate the constant, we obtain the induced constraint 
\begin{eqnarray}
G(x_1, x_2, x_3)= 2x_{3}^{2}-2x_{1}x_{2}=0.
\label{Ex1-G}
\end{eqnarray}

As an example of the dynamics of the Nambu triplet in Eq. (\ref{Ex1}),
consider a one-dimensional harmonic oscillator whose Hamiltonian is given by
\begin{eqnarray}
H(q,p)=\frac{1}{2m}p^{2}+\frac{m\omega^{2}}{2}q^{2}.
\label{Ex1-H}
\end{eqnarray}
The Hamilton equations of motion for the triplet are as follows:
\begin{eqnarray}
\frac{d}{dt}q^{2}=\frac{2}{m}qp,~~~~~~
\frac{d}{dt}p^{2}=-2m\omega^{2}qp,~~~~~~
\frac{d}{dt}(qp)=\frac{1}{m}p^{2}-m\omega^{2}q^{2}.
\label{Ex1-H-eqs}
\end{eqnarray}
Let us derive these equations from the $N=3$ Nambu equations with two Nambu Hamiltonians $(F,G)$.
One of the Hamiltonians, $F$, is equal to the original Hamiltonian $H(q,p)$, Eq. (\ref{Ex1-H}),
\begin{eqnarray}
F(x_1, x_2, x_3)=H(q,p)=\frac{1}{2m}x_{2}+\frac{m\omega^{2}}{2}x_{1},
\label{Ex1-F}
\end{eqnarray}
and the other Hamiltonian, $G$, is given by the induced constraint, Eq. (\ref{Ex1-G}).
The $N=3$ Nambu equations are
\begin{eqnarray}
\frac{d}{dt}x_{i}=\frac{\partial(x_{i},F,G)}{\partial(x_{1},x_{2},x_{3})},
\label{Ex1-HN-eq}
\end{eqnarray}
and each equation is given by
\begin{eqnarray}
\frac{d}{dt}x_{1}=\frac{2}{m}x_{3},~~~~~~
\frac{d}{dt}x_{2}=-2m\omega^{2}x_{3},~~~~~~
\frac{d}{dt}x_{3}=\frac{1}{m}x_{2}-m\omega^{2}x_{1}.
\label{Ex1-HN-eqs}
\end{eqnarray}
These equations are equivalent to the Hamilton equations of motion in Eq. (\ref{Ex1-H-eqs}).
\\

{\bf (b) $N=4$: classical nonlinear systems}\\
Consider four variables, two of them being composites,
\begin{eqnarray}
x_1=q,~~~ x_2=p,~~~ x_3=q^{2},~~~ x_4=p^{2},
\label{Ex2}
\end{eqnarray}
which satisfy the following relations:
\begin{eqnarray}
&& \{x_1, x_2\}_{\mbox{\tiny{PB}}}=1,~~~~~~
~~ \{x_1, x_3\}_{\mbox{\tiny{PB}}}=0,~~
~~ \{x_1, x_4\}_{\mbox{\tiny{PB}}}=2x_{2},
\nonumber \\
&& \{x_2, x_3\}_{\mbox{\tiny{PB}}}=-2x_{1},~
~~ \{x_2, x_4\}_{\mbox{\tiny{PB}}}=0,~~
~~ \{x_3, x_4\}_{\mbox{\tiny{PB}}}=4x_{1}x_{2}.
\label{Ex2-PB}
\end{eqnarray}
Then, the conditions in Eq. (\ref{condition}) become
\begin{eqnarray}
&& \frac{\partial (G_{1},G_{2})}{\partial (x_{1},x_{2})}=4x_{1}x_{2},~
~~ \frac{\partial (G_{1},G_{2})}{\partial (x_{1},x_{3})}=0,~~
~~ \frac{\partial (G_{1},G_{2})}{\partial (x_{1},x_{4})}=-2x_{1},
\nonumber \\
&& \frac{\partial (G_{1},G_{2})}{\partial (x_{2},x_{3})}=2x_{2},~~~~
~~ \frac{\partial (G_{1},G_{2})}{\partial (x_{2},x_{4})}=0,~~
~~ \frac{\partial (G_{1},G_{2})}{\partial (x_{3},x_{4})}=1,
\label{Ex2-dG}
\end{eqnarray}
and $G_{1}$ and $G_{2}$ are given by
$G_{1}= x_{3}-x_{1}^{2}+C_{1}$ and $G_{2}= x_{4}-x_{2}^{2}+C_{2}$,
where $C_{1}$ and $C_{2}$ are constants.
By redefining $G_1$ and $G_2$, we obtain the induced constraints 
\begin{eqnarray}
&&G_{1}= x_{3}-x_{1}^{2}=0,
\label{Ex2-G1}\\
&&G_{2}= x_{4}-x_{2}^{2}=0.
\label{Ex2-G2}
\end{eqnarray}

As an example of the dynamics of the Nambu quartet in Eq. (\ref{Ex2}),
consider a one-dimensional nonlinear system whose Hamiltonian is given by
\begin{eqnarray}
H(q,p)=\frac{1}{2m}p^{2}+V(q),
\label{Ex2-H}
\end{eqnarray}
where $V(q)$ is an anharmonic potential.
The Hamilton equations of motion for the quartet are written as follows:
\begin{eqnarray}
\frac{d}{dt}q=\frac{1}{m}p,~~~~~~
\frac{d}{dt}p=-\frac{\partial V}{\partial q},~~~~~~
\frac{d}{dt}q^{2}=\frac{2}{m}qp,~~~~~~
\frac{d}{dt}p^{2}=-2p\frac{\partial V}{\partial q}.
\label{Ex2-H-eqs}
\end{eqnarray}
Let us derive these equations from the $N=4$ Nambu equations with three Nambu Hamiltonians $(F,G_{1},G_{2})$.
One of the Hamiltonians, $F$, is equal to the original Hamiltonian $H(q,p)$,
\begin{eqnarray}
F(x_1, x_2, x_3,x_4)=H(q,p)=\frac{1}{2m}x_{4}+\tilde{V}(x_{1},x_{3}),
\label{Ex2-F}
\end{eqnarray}
where $\tilde{V}(x_{1},x_{3})=V(q)$.
The other two Hamiltonians, $G_{1}$ and $G_{2}$, are given 
by the induced constraints in Eqs. (\ref{Ex2-G1}) and (\ref{Ex2-G2}).
The $N=4$ Nambu equations are
\begin{eqnarray}
\frac{d}{dt}x_{i}=
\frac{\partial(x_{i},F,G_{1},G_{2})}{\partial(x_{1},x_{2},x_{3},x_{4})},
\label{Ex2-HN-eq}
\end{eqnarray}
and each equation is given by
\begin{eqnarray}
&&\frac{d}{dt}x_{1}=\frac{1}{m}x_{2},~~~~~~~~~~~
\frac{d}{dt}x_{2}=-\frac{\partial \tilde{V}}{\partial x_{1}}
-2x_{1}\frac{\partial \tilde{V}}{\partial x_{3}},\nonumber\\
&&\frac{d}{dt}x_{3}=\frac{2}{m}x_{1}x_{2},~~~~~~~~
\frac{d}{dt}x_{4}=-2x_{2}\frac{\partial \tilde{V}}{\partial x_{1}}
-4x_{1}x_{2}\frac{\partial \tilde{V}}{\partial x_{3}}.
\label{Ex2-HN-eqs}
\end{eqnarray}
These equations are equivalent to the Hamilton equations of motion in Eq. (\ref{Ex2-H-eqs}).
\\

{\bf Some comments}\\
Here we make some comments on the functional forms of Nambu Hamiltonians.
In some cases, the functional form of $(G_1, ..., G_{N-2})$ cannot be determined uniquely.
For example, for $(x_1, x_2, x_3, x_4)=(q, p, q^2, q^3)$,
one of the Poisson brackets in the consistency condition in Eq. (\ref{condition}) is given by
$\{x_{2}, x_{4}\}_{\mbox{\tiny{PB}}}=\{p, q^3\}_{\mbox{\tiny{PB}}}=-3q^2$.
This can be expressed as either $\{x_{2}, x_{4}\}_{\mbox{\tiny{PB}}}=-3x_1^2$ or $-3x_3$,
which leads to $G_2=x_4-x_1^3$ or $x_4-3x_3x_1+2x_1^3$, respectively.
Although we can choose either expression in classical mechanics,
we must choose the latter expression, $\{x_{2}, x_{4}\}_{\mbox{\tiny{PB}}}=-3x_3$,
in quantum or semiclassical mechanics.
As shown in the next section,
we must express the Poisson brackets in the consistency condition of Eq. (\ref{condition}) 
using variables of the highest order possible.

Also, in some cases we cannot uniquely determine the functional form of the Hamiltonian $F$ in classical mechanics.
In the next section we present a prescription to determine the functional form of $F$
in quantum or semiclassical systems.

\section{Hidden Nambu mechanics in quantum/semiclassical systems}
\label{Hidden Nambu mechanics in quantum/semiclassical systems}
\subsection{Quantum/semiclassical dynamics}
\label{Quantum/semiclassical dynamics}

Consider a quantum system of a doublet $(\hat{q},\hat{p})$ with a Hamiltonian operator
\begin{eqnarray}
\hat{H}=H(\hat{q},\hat{p})=\frac{1}{2m}\hat{p}^{2}+V(\hat{q}).
\label{Q-H}
\end{eqnarray}
The dynamics of a quantum operator $\hat{A}=A(\hat{q},\hat{p})$ is given by the Heisenberg equation,
\begin{eqnarray}
\frac{d}{dt}\hat{A}(t)=\frac{1}{i\hbar}\left[\hat{A}(t),\hat{H}\right]
=\frac{1}{i\hbar}\left(\hat{A}(t)\hat{H}-\hat{H}\hat{A}(t)\right).
\label{He-eq}
\end{eqnarray}
In this work we focus on the dynamics of the expectation value of the quantum operator,
$\langle\hat{A}(t)\rangle = \langle \psi|\hat{A}(t)|\psi\rangle$,
where $|\psi\rangle$ is a quantum state. 
The time evolution of $\langle\hat{A}(t)\rangle$ is given by taking the expectation value of
both sides of the Heisenberg equation,
\begin{eqnarray}
\frac{d}{dt}\langle\hat{A}(t)\rangle=\frac{1}{i\hbar}\langle\left[\hat{A}(t),\hat{H}\right]\rangle.
\label{EHe-eq}
\end{eqnarray}
This equation gives the exact quantum dynamics, and we can consider several approximated dynamics.
The lowest-order approximation is simply the classical Hamiltonian dynamics,
\begin{eqnarray}
\frac{d}{dt}\langle\hat{q}\rangle
=\frac{\partial}{\partial \langle\hat{p}\rangle}H(\langle\hat{q}\rangle,\langle\hat{p}\rangle),~~~~~~
\frac{d}{dt}\langle\hat{p}\rangle
=-\frac{\partial}{\partial \langle\hat{q}\rangle}H(\langle\hat{q}\rangle,\langle\hat{p}\rangle).
\label{LOeqs}
\end{eqnarray}
A systematic approximation scheme to derive higher-order semiclassical dynamics, 
the quantized Hamiltonian dynamics \cite{PrezhdoPereverzev,Prezhdo}, 
has been developed.

In quantum or semiclassical systems there might exist conserved quantities other than 
$\langle\hat{H}\rangle=\langle H(\hat{q},\hat{p})\rangle$,
the expectation value of the original Hamiltonian.
Moreover, in some cases such conserved quantities might be identified as
the constraints $G_{c}$ in the hidden Nambu mechanics.
This means that the Nambu structure could be hidden in quantum or semiclassical systems with
nontrivial constraints $G_{c}\ne 0$, which are trivial ($G_{c}=0$) in classical systems.

\subsection{How to find the hidden Nambu structure}
\label{How to find the hidden Nambu structure}

The procedure to find the Nambu structure hidden in quantum or semiclassical systems is as follows.
\\

\underline {Step (1)}:~Start from a classical system of a canonical doublet $(q,p)$ with a Hamiltonian $H(q,p)$. 
Choose a Nambu $N$-plet $(x_1, x_2, ..., x_{N})$ and 
determine the functional form of $N-2$ trivial constraints $(G_1, G_2, ..., G_{N-2})$
by means of the consistency conditions in Eq. (\ref{condition}),
where the Poisson brackets are represented in terms of variables of the highest order possible.
\\

\underline {Step (2)}:~Consider a quantum system of a doublet $(\hat{q},\hat{p})$ 
with a Hamiltonian $\hat{H}=H(\hat{q},\hat{p})$, Eq. (\ref{Q-H}), 
which corresponds to the classical Hamiltonian $H(q,p)$.
Replace the Nambu $N$-plet with the corresponding expectation values of quantum operators. 
For example, $x_{1}=q^{2} \to \langle\hat{q}^{2}\rangle$. 
\\

\underline {Step (3)}:~Determine the functional form of $F(x_1, x_2,..., x_N)$ by representing $\langle\hat{H}\rangle$
as a function of the Nambu $N$-plet. 
If $\langle\hat{H}\rangle$ includes an expectation value 
$\langle\hat{O}\rangle$ which is not a member of the Nambu $N$-plet, 
we reduce $\langle\hat{O}\rangle$ to a function of the Nambu $N$-plet
by means of the {\it zero-cumulant approximation} that ignores the cumulant, 
\begin{eqnarray}
\langle\hat{O}\rangle_{c}\simeq 0. 
\label{zca}
\end{eqnarray} 
For example, for 
$(x_1, x_2, x_3)=(\langle\hat{q}\rangle,\langle\hat{p}\rangle, \langle\hat{q}^{2}\rangle)$,
if $\langle\hat{H}\rangle$ includes $\langle\hat{q}^{4}\rangle$, 
it is approximated as $\langle\hat{q}^{4}\rangle\simeq 3\langle\hat{q}^{2}\rangle^{2}-2\langle\hat{q}\rangle^{4}
=3x_{3}^{2}-2x_{1}^{4}$ by means of $\langle\hat{q}^{4}\rangle_{c}\simeq 0$ followed by
$\langle\hat{q}^{3}\rangle_{c}\simeq 0$.
\\

\underline {Step (4)}:~The other Nambu Hamiltonians $G_{c}$ $(c=1, ..., N-2)$ 
are given by the same functional forms as the trivial constraints. 
They are in general nontrivial, $G_{c}\ne 0$,  because of quantum fluctuation.
\\

\underline {Step (5)}:~If the quantities $(F, G_1,..., G_{N-2})$ are all conserved
in quantum or some semiclassical dynamics, 
the dynamics of the Nambu $N$-plet can be cast into the Nambu form in Eq. (\ref{HN-eq}).
\\
 
The zero-cumulant approximation is similar to the approximation adopted 
in the quantized Hamiltonian dynamics \cite{PrezhdoPereverzev}.
However, it is not the only approximation for the Hamiltonian $F$ in the hidden Nambu mechanics.
It is also possible to consider an approximation that ignores the quantum fluctuation, for example
$\langle(\hat{q}-\langle \hat{q}\rangle)^{n}\rangle\simeq 0$.  
As for the example shown in Step (3), this approximation leads to 
$\langle\hat{q}^{4}\rangle\simeq 
6\langle\hat{q}^{2}\rangle\langle\hat{q}\rangle^{2}-5\langle\hat{q}\rangle^{4}
=6x_{3}x_{1}^{2}-5x_{1}^{4}$.

The resulting Nambu equations for quantum/semiclassical systems are the same 
as the ones for classical systems, and 
quantum properties are introduced through nonzero constraints $G_{c}\ne 0$ $(c=1, ..., N-2)$. 
This might imply that the replacement 
\begin{eqnarray}
G_{c}=0~\to~ G_{c}=C_{c},
\label{quantization}
\end{eqnarray}
where $C_{c}$ are nonzero constants, could be regarded as a kind of ``quantization" scheme
for the Nambu mechanics.
However, as opposed to various attempts to quantize the Nambu mechanics 
proposed so far \cite{DitoFlatoSternheimerTakhtajan,AwataLiMinicYoneya,CurtrightZachos,AxenidesFloratosNicolis,HoMatsuo},
this replacement only gives a scheme to quantize the {\it  hidden} Nambu mechanics.
Furthermore, this replacement is incomplete even as a quantization scheme for 
the hidden Nambu mechanics,
because the constants $C_{c}$ in general depend on the models and the initial conditions.
Therefore the procedure presented here is {\it not} for quantizing the Nambu mechanics,
but just for finding the hidden Nambu structures in quantum/semiclassical systems.

The resulting hidden Nambu mechanics is a volume-preserving dynamics of the expectation values 
of the quantum operators. 
If the Nambu multiplet includes the variables $(\langle\hat{q}\rangle,\langle\hat{p}\rangle)$,
the Nambu mechanics can be regarded as a kind of 
quantized Hamiltonian dynamics \cite{PrezhdoPereverzev}.
In some cases, such Nambu mechanics can be reduced to 
the effective Hamiltonian dynamics by explicitly solving the constraints $G_{c}\ne 0$.
In the next subsection we will see an example of such reduction.

\subsection{Examples}
\label{Examples-q}

Here we present two examples; one is an example of exact quantum dynamics and the other is semiclassical. 
In the latter example, the resulting Nambu mechanics can be reduced to the Hamiltonian dynamics 
with the effective Hamiltonian.  
They correspond to the examples shown in Sect. \ref{Examples}, and therefore
we will show the procedure after Step (2).  
 \\

{\bf (a) $N=3$: quantum harmonic oscillator}\\
Consider three expectation values which correspond to the Nambu triplet in Eq. (\ref{Ex1}),
\begin{eqnarray}
x_{1}=\langle\hat{q}^2\rangle,~~ 
x_{2}=\langle\hat{p}^2\rangle,~~ 
x_{3}=\langle(\hat{q}\hat{p})_{s}\rangle=\langle\left(\frac{\hat{q}\hat{p}+\hat{p}\hat{q}}{2}\right)\rangle.
\label{Ex1q}
\end{eqnarray}
The quantum Hamiltonian of a one-dimensional harmonic oscillator is given by
\begin{eqnarray}
\hat{H}=H(\hat{q},\hat{p})=\frac{1}{2m}\hat{p}^{2}+\frac{m\omega^{2}}{2}\hat{q}^{2}.
\label{Ex1-Hq}
\end{eqnarray}
Then, one of the Nambu Hamiltonians, $F$, can be obtained without any approximation,
\begin{eqnarray}
F(x_1, x_2, x_3)=\langle\hat{H}\rangle
=\frac{1}{2m}\langle\hat{p}^{2}\rangle+\frac{m\omega^{2}}{2}\langle\hat{q}^{2}\rangle
=\frac{1}{2m}x_{2}+\frac{m\omega^{2}}{2}x_{1}.
\label{Ex1-Fq}
\end{eqnarray}
The other Nambu Hamiltonian, $G$, is given by the same functional form as Eq. (\ref{Ex1-G}),
\begin{eqnarray}
G(x_1, x_2, x_3)= 2x_{3}^{2}-2x_{1}x_{2},
\label{Ex1-Gq}
\end{eqnarray}
which is nonzero in general due to quantum fluctuation.
We can see that both $F$ and $G$ are conserved in 
the exact quantum dynamics,
\begin{eqnarray}
\frac{d}{dt}\langle\hat{q}^{2}\rangle=\frac{2}{m}\langle(\hat{q}\hat{p})_{s}\rangle,~~~
\frac{d}{dt}\langle\hat{p}^{2}\rangle=-2m\omega^{2}\langle(\hat{q}\hat{p})_{s}\rangle,~~~
\frac{d}{dt}\langle(\hat{q}\hat{p})_{s}\rangle
=\frac{1}{m}\langle\hat{p}^{2}\rangle-m\omega^{2}\langle\hat{q}^{2}\rangle.
\label{Ex1-Hq-eqs}
\end{eqnarray}
The Nambu equations of Eq. (\ref{Ex1-HN-eqs}) are equivalent to 
these exact equations; that is,
the Nambu structure is hidden in the exact quantum dynamics of a harmonic oscillator.
\\

{\bf (b) $N=4$: semiclassical nonlinear systems}\\
Consider four expectation values which correspond to the Nambu quartet in Eq. (\ref{Ex2}),
\begin{eqnarray}
x_1=\langle\hat{q}\rangle,~~ x_2=\langle\hat{p}\rangle,
~~ x_3=\langle\hat{q}^{2}\rangle,~~ x_4=\langle\hat{p}^{2}\rangle.
\label{Ex2q}
\end{eqnarray}
The quantum Hamiltonian of a one-dimensional nonlinear system is given by 
$\hat{H}=H(\hat{q},\hat{p})$, Eq. (\ref{Q-H}), with an anharmonic potential $\hat{V}=V(\hat{q})$.
The Nambu Hamiltonian $F$ can be obtained as an approximation of 
$\langle\hat{H}\rangle$,
\begin{eqnarray}
\langle\hat{H}\rangle
=\frac{1}{2m}\langle\hat{p}^{2}\rangle+\langle V(\hat{q})\rangle
\simeq\frac{1}{2m}x_{4}+\tilde{V}(x_{1},x_{3})=F(x_1, x_2, x_3,x_4),
\label{Ex2-Fq}
\end{eqnarray}
where the functional form of the reduced potential $\tilde{V}(x_{1},x_{3})$ 
is uniquely determined by means of the zero-cumulant approximation, Eq. (\ref{zca}). 
The other Nambu Hamiltonians are given by the same functional forms as Eqs. (\ref{Ex2-G1}) and (\ref{Ex2-G2}),
$G_{1}= x_{3}-x_{1}^{2}$ and $G_{2}= x_{4}-x_{2}^{2}$,
which are nonzero in general due to quantum fluctuation.
We can see that all of $F$, $G_{1}$, and $G_{2}$ are conserved in 
the following approximated dynamics:
\begin{eqnarray}
&&~\frac{d}{dt}\langle\hat{q}\rangle=\frac{1}{m}\langle\hat{p}\rangle,~~~~~~~~~~~~~~~~~~~~~~~~~~~
\frac{d}{dt}\langle\hat{p}\rangle=
\langle-\frac{\partial \hat{V}}{\partial q}\rangle
=f\simeq \tilde{f}(\langle\hat{q}\rangle,\langle\hat{q}^{2}\rangle),\nonumber\\
&&\frac{d}{dt}\langle\hat{q}^{2}\rangle=\frac{2}{m}\langle(\hat{q}\hat{p})_{s}\rangle
\simeq\frac{2}{m}\langle\hat{q}\rangle\langle\hat{p}\rangle,~~~~~~
\frac{d}{dt}\langle\hat{p}^{2}\rangle=2\langle(-\frac{\partial \hat{V}}{\partial q}\hat{p})_{s}\rangle
\simeq 2\tilde{f}(\langle\hat{q}\rangle,\langle\hat{q}^{2}\rangle)\langle\hat{p}\rangle,
\label{Ex2-Hq-eqs}
\end{eqnarray}
where $\tilde{f}$ is determined by the zero-cumulant approximation in Eq. (\ref{zca}) if necessary.
This is a semiclassical dynamics which corresponds to the lowest order of 
the quantized Hamiltonian dynamics \cite{Prezhdo}.
We can also see that the $N=4$ Nambu equations of Eq. (\ref{Ex2-HN-eqs}) are equivalent to 
these semiclassical equations; that is,
the Nambu structure is hidden in the semiclassical nonlinear dynamics.

This semiclassical dynamics can be regarded as 
the frozen Gaussian wave packet dynamics \cite{Heller}, where the quantum wave function is approximated by 
a Gaussian wave packet with a constant width $\sigma$,
\begin{eqnarray}
\psi_{\rm FG}(q,t)=\langle q|\psi_{\rm FG}(t)\rangle=\left(\frac{1}{2\pi\sigma^{2}}\right)^{\frac{1}{4}}
{\rm exp}\left[-\frac{1}{4\sigma^{2}}(q-q_{c}(t))^{2}+\frac{i}{\hbar}p_{c}(t)(q-q_{c}(t))\right].
\label{Ex2-FGWP}
\end{eqnarray}
Here $q_{c}$ is the center of the wave packet and $p_{c}$ is that in momentum space.
The time evolution of the variables $(q_{c},p_{c})$ can be determined 
by means of the time-dependent variational principle \cite{FeldmeierSchnack},
\begin{eqnarray}
\delta\int^{t_{\rm f}}_{t_{\rm i}}dt~
\langle \psi_{\rm FG}(t)|\left(i\hbar\frac{d}{dt}-\hat{H}\right)|\psi_{\rm FG}(t)\rangle=0,
\label{TDVP}
\end{eqnarray}
by taking the frozen Gaussian wave function of Eq. (\ref{Ex2-FGWP}) as a trial function.
The resulting variational equations are semiclassical equations which have the same forms as 
the Hamilton equations of motion in Eq. (\ref{H-eq}),
\begin{eqnarray}
\frac{d}{dt}q_{c}=\frac{\partial H_{c}}{\partial p_{c}},~~~~~~
\frac{d}{dt}p_{c}=-\frac{\partial H_{c}}{\partial q_{c}}.
\label{TDVPeqs}
\end{eqnarray}
Here $H_{c}(q_{c},p_{c})=\langle \psi_{\rm FG}|\hat{H}|\psi_{\rm FG}\rangle$ is the effective Hamiltonian
modified by the quantum correction.

Evaluating the expectation values in Eq. (\ref{Ex2q})
by means of the state $|\psi_{\rm FG}(t)\rangle$, we obtain
$x_{1}=\langle\hat{q}(t)\rangle=\langle\psi_{\rm FG}(t)|\hat{q}|\psi_{\rm FG}(t)\rangle=q_{c}$, 
$x_{2}=p_{c}$, $x_{3}=q_{c}^{2}-\sigma^{2}$, and   
$x_{4}=p_{c}^{2}-\hbar^{2}/(4\sigma^{2})$.
Then, $G_{1}$ and $G_{2}$ are given by
\begin{eqnarray}
&&G_{1}= x_{3}-x_{1}^{2}=\sigma^{2},
\label{Ex2-G1qFG}\\
&&G_{2}= x_{4}-x_{2}^{2}=\frac{\hbar^{2}}{4\sigma^{2}}.
\label{Ex2-G2qFG}
\end{eqnarray}
Using these nontrivial constraints,
we can show that the Nambu Hamiltonian $F$ in Eq. (\ref{Ex2-Fq}) is equivalent to $H_{c}(q_{c},p_{c})$.
This is because the zero-cumulant approximation for the Nambu quartet in Eq. (\ref{Ex2q})
is exact in the case of the frozen Gaussian wave packet.
Using Eqs. (\ref{Ex2-G1qFG}) and (\ref{Ex2-G2qFG}), 
we can also show that the Nambu equations of Eq. (\ref{Ex2-HN-eqs}) are reduced to 
the variational equations of Eq. (\ref{TDVPeqs}).
That is, the Nambu structure is hidden in the semiclassical dynamics of the frozen Gaussian wave packet.
In Sect. 5.1 we present a numerical demonstration of the semiclassical tunneling dynamics in a metastable system.
\\

\section{Many-degrees-of-freedom extension}
\label{Many-degrees-of-freedom extension}
It is straightforward to extend our formalism to many-degrees-of-freedom systems.
However, the resulting classical or quantum/semiclassical hidden Nambu mechanics becomes pathological,
because the Nambu bracket itself has a serious problem in interacting systems \cite{Nambu,Takhtajan,HoMatsuo}.
\subsection{Difficulties in the Nambu bracket}
\label{Difficulties in the Nambu bracket}

Consider a Hamiltonian system of $n$ canonical doublets $(q^{(1)},p^{(1)}, ..., q^{(n)},p^{(n)})$. 
The Hamilton equation of motion can be written in the same form as Eq. (\ref{H-eq}) 
by extending the definition of the Poisson bracket in Eq. (\ref{PB}),
\begin{eqnarray}
\{A, B\}_{\mbox{\tiny{PB}}} \equiv 
\sum_{{\alpha}=1}^{n}\frac{\partial (A,B)}{\partial (q^{(\alpha)},p^{(\alpha)})}=
\sum_{{\alpha}=1}^{n} \left(
\frac{\partial A}{\partial q^{(\alpha)}}\frac{\partial B}{\partial p^{(\alpha)}}
 - \frac{\partial A}{\partial p^{(\alpha)}}\frac{\partial B}{\partial q^{(\alpha)}}\right),
\label{PBn}
\end{eqnarray}  
where $A$ and $B$ are any functions of the $2n$ variables.
Since the dynamics is divergenceless,
\begin{eqnarray}
\frac{\partial }{\partial q^{(\alpha)}}\frac{d}{dt}q^{(\alpha)}+ 
\frac{\partial }{\partial p^{(\alpha)}}\frac{d}{dt}p^{(\alpha)}=
\frac{\partial }{\partial q^{(\alpha)}}\{q^{(\alpha)},H \}_{\mbox{\tiny{PB}}}+
\frac{\partial }{\partial p^{(\alpha)}}\{p^{(\alpha)},H \}_{\mbox{\tiny{PB}}}=0,
\label{divHn}
\end{eqnarray}
the Liouville theorem holds.
The Poisson bracket in Eq. (\ref{PBn}) satisfies the Jacobi identity of Eq. (\ref{JI})
and therefore we can define canonical transformations of the $2n$ variables 
in a consistent manner.

On the other hand, the Nambu mechanics has a problem in the many-degrees-of-freedom extension.
Consider a system of $n$ Nambu $N$-plets 
$(x_1^{(1)}, ..., x_N^{(1)}, ..., x_1^{(n)}, ..., x_N^{(n)})$. 
The time evolution of the $N\times n$ variables can be given in the same form as Eq. (\ref{N-eq})
by extending the definition of the Nambu bracket in Eq. (\ref{NB}),
\begin{eqnarray}
\{{A}_1, {A}_2, ..., {A}_N\}_{\mbox{\tiny{NB}}} 
&\equiv& \sum_{{\alpha}=1}^{n}
\frac{\partial ({A}_1, {A}_2, ..., {A}_N)}{\partial (x_1^{(\alpha)}, x_2^{(\alpha)}, ..., x_N^{(\alpha)})}
\nonumber \\
&=& \sum_{{\alpha}=1}^{n} ~\sum_{i_1, i_2, ..., i_N=1}^{N} \varepsilon_{i_1 i_2 \cdots i_N}
\frac{\partial {A}_1}{\partial x_{i_1}^{(\alpha)}}\frac{\partial {A}_2}{\partial x_{i_2}^{(\alpha)}}
 ... \frac{\partial {A}_{N}}{\partial x_{i_N}^{(\alpha)}},
\label{NBn}
\end{eqnarray}
where ${A}_{a}$ $(a=1, ..., N)$ are any functions of the $N\times n$ variables.
Because of the divergenceless property,
\begin{eqnarray}
\sum_{i=1}^{N}\frac{\partial }{\partial x_{i}^{(\alpha)}}\frac{d}{dt}x_{i}^{(\alpha)} =
\sum_{i=1}^{N}\frac{\partial }{\partial x_{i}^{(\alpha)}}
\{x_{i}^{(\alpha)}, {H}_1, ..., {H}_{N-1}\}_{\mbox{\tiny{NB}}}=0,
\label{divNn}
\end{eqnarray}
the Liouville theorem holds.
The Nambu bracket of Eq. (\ref{NBn}), however, fails to satisfy the fundamental identity in Eq. (\ref{FI}) 
if the $N$-plets interact with each other \cite{Takhtajan,HoMatsuo}.
Therefore we cannot define consistent canonical transformations 
of the $N\times n$ variables in general \cite{Nambu}. 

\subsection{Hidden Nambu mechanics in many-degrees-of-freedom systems}
\label{Hidden Nambu mechanics in many-degrees-of-freedom systems}

Although the Nambu bracket has a serious problem in many-degrees-of-freedom systems,
it is still possible to extend our hidden Nambu formalism to such systems.
The resulting hidden Nambu mechanics is {\it the Nambu mechanics without the fundamental identity}.

We start from a Hamiltonian system of $n$ canonical doublets with a Hamiltonian 
$H=H(q^{(1)},p^{(1)}, ..., q^{(n)},p^{(n)})$.
Then we introduce $N$ variables $x_i^{(\alpha)}=x_i^{(\alpha)}(q^{(\alpha)}, p^{(\alpha)})$ 
$(i=1, ..., N)$ for each $\alpha$.
We assume that at least $N-1$ of 
$\partial(x_{i}^{(\alpha)}, x_{j}^{(\alpha)})/\partial(q^{(\alpha)}, p^{(\alpha)})$ 
do not vanish for each $\alpha$.
In this case, the time evolution of any function 
$\tilde{f}(x_1^{(1)}, ..., x_N^{(1)}, ..., x_1^{(n)}, ..., x_N^{(n)}) = 
f(q^{(1)},p^{(1)}, ..., q^{(n)},p^{(n)})$
can be written via the Hamilton equation of motion, 
\begin{eqnarray}
\frac{d \tilde{f}}{dt} = \frac{d f}{dt} = \sum_{\alpha =1}^{n}\frac{\partial(f, H)}{\partial(q^{(\alpha)}, p^{(\alpha)})} 
= \frac{1}{2} \sum_{\alpha =1}^{n}\sum_{i, j=1}^{N} 
\frac{\partial (\tilde{f}, F)}{\partial (x_{i}^{(\alpha)}, x_{j}^{(\alpha)})}
\frac{\partial(x_{i}^{(\alpha)}, x_{j}^{(\alpha)})}{\partial(q^{(\alpha)}, p^{(\alpha)})},
\label{H-eq2n}
\end{eqnarray}
where $F(x_1^{(1)}, ..., x_N^{(1)}, ..., x_1^{(n)}, ..., x_N^{(n)}) = H(q^{(1)},p^{(1)}, ..., q^{(n)},p^{(n)})$.
We introduce
$N-2$ functions $G_{c}^{(\alpha)}=G_{c}^{(\alpha)}(x_1^{(\alpha)}, ..., x_N^{(\alpha)})$ $(c=1, ..., N-2)$
that satisfy the consistency conditions for each $\alpha$,
\begin{eqnarray}
\frac{1}{(N-2)!} \sum_{k_1, ..., k_{N-2}=1}^{N}
\varepsilon_{i j k_1 \cdots k_{N-2}}
\frac{\partial (G_1^{(\alpha)}, ..., G_{N-2}^{(\alpha)})}
{\partial (x_{k_1}^{(\alpha)}, ..., x_{k_{N-2}}^{(\alpha)})}=
\frac{\partial(x_{i}^{(\alpha)}, x_{j}^{(\alpha)})}{\partial(q^{(\alpha)}, p^{(\alpha)})},
\label{condition-n}
\end{eqnarray}
then Eq. (\ref{H-eq2n}) can be rewritten in the same form as Eq. (\ref{N-eq}),
\begin{eqnarray}
\frac{d \tilde{f}}{dt} = 
\{\tilde{f}, F, G_1, ..., G_{N-2}\}_{\mbox{\tiny{NB}}},
\label{HN-eqn}
\end{eqnarray}
where the Hamiltonians $G_c$ are defined as the sum of each $G_c^{(\alpha)}$,
\begin{eqnarray}
G_{c}(x_1^{(1)}, ..., x_N^{(1)}, ..., x_1^{(n)}, ..., x_N^{(n)})=
\sum_{\alpha =1}^{n} G_{c}^{(\alpha)}(x_1^{(\alpha)}, ..., x_N^{(\alpha)}).
\label{Gn}
\end{eqnarray}
The Nambu mechanics is hidden in classical many-degrees-of-freedom systems.
The Liouville theorem holds in the hidden mechanics, though
the fundamental identity does not hold.
Such hidden Nambu mechanics is an example of dynamics without the canonical structure 
\cite{SatoYoshida}.

Taking the same procedure presented in Sect. 3.2,
we can find the Nambu structure in quantum/semiclassical many-degrees-of-freedom systems:\\

\underline {Step (1)}:~Start from the classical hidden Nambu mechanics shown above.\\

\underline {Step (2)}:~Replace the Nambu $N$-plets with the corresponding expectation values of quantum operators. 
For example, $x_{1}=q^{(1)}q^{(2)} \to \langle\hat{q}^{(1)}\hat{q}^{(2)}\rangle$. \\

\underline {Step (3)}:~Determine the functional form of $F$. Use the zero-cumulant approximation if necessary.\\

\underline {Step (4)}:~$(G_1,..., G_{N-2})$ are given by the same functional forms as the classical forms. \\

\underline {Step (5)}:~If $(F, G_1,..., G_{N-2})$ are all conserved, 
the dynamics of the Nambu $N$-plets can be cast into the Nambu form in Eq. (\ref{HN-eqn}).\\

In Sect. 5.2 we present a numerical demonstration of the semiclassical dynamics 
of a two-body system.
\\

\section{Numerical results for semiclassical dynamics}
\label{Numerical results for semiclassical dynamics}

Here we give two numerical results for $N=4$ hidden Nambu mechanics
equivalent to the semiclassical frozen Gaussian wave packet dynamics
in one- and two-degrees-of-freedom systems.
We choose the same systems as used in the applications of 
the quantized Hamiltonian dynamics \cite{PrezhdoPereverzev}.
We compare the results with the corresponding quantum and classical results.
In both systems, the time developments in the Nambu and classical mechanics are numerically evaluated by
using the fourth-order Runge--Kutta integrator,\footnote{
Similar to the symplectic integrator for Hamiltonian dynamics,
we can formulate some volume-preserving integrators for Nambu mechanics
(see Ref.\cite{Modin}; A.~Horikoshi, in preparation).
}
while the propagation of the quantum wave function is numerically evaluated by
a split-operator method, 
which is a hybrid of Cayley's form and the Suzuki--Trotter decomposition \cite{WatanabeTsukada}.
As the initial wave function for the quantum dynamics, we take the Gaussian wave packet 
$\psi_{\rm FG}(q,0)$, Eq. (\ref{Ex2-FGWP}).
The initial conditions for the Nambu mechanics are given by the expectation values of the quantum operators 
with respect to that initial state,
and those for the classical mechanics are given by the center of the initial wave packet $(q_{c}(0),p_{c}(0))$.
We choose the width of the initial wave packet as $\sigma=\sqrt{\hbar/(2m\omega)}$,
for which the frozen Gaussian wave packet dynamics becomes exact for a harmonic oscillator. 

\subsection{Metastable cubic potential}
\label{Metastable cubic potential}

\begin{figure}[htbp]
\hspace{-10mm}
 \begin{minipage}{0.5\hsize}
  \begin{center}
  \includegraphics[width=70mm]{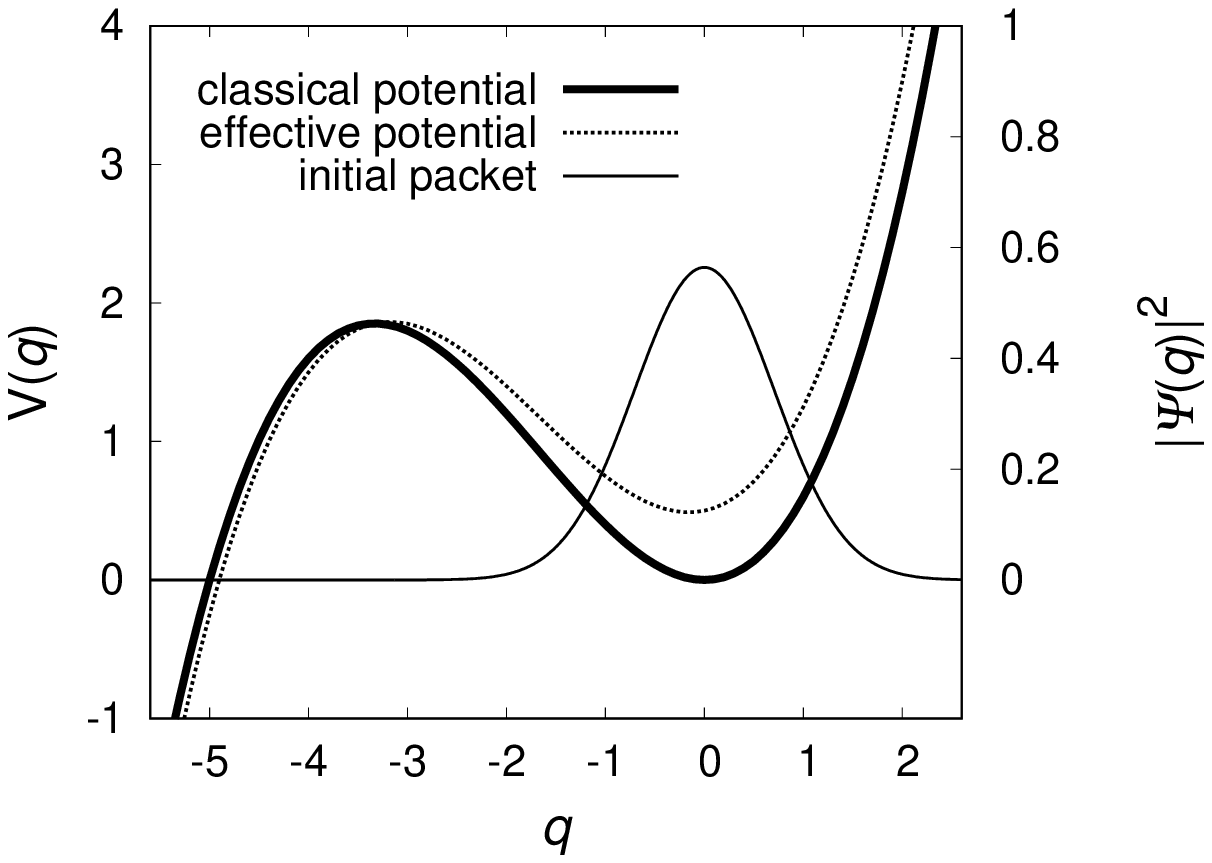}
  \vspace{0mm}
  \hspace{-15mm}
  \subcaption{Potentials and the initial packet.}
  \label{fig1a}
  \end{center}
 \end{minipage}
 \begin{minipage}{0.5\hsize}
  \begin{center}
  \includegraphics[width=70mm]{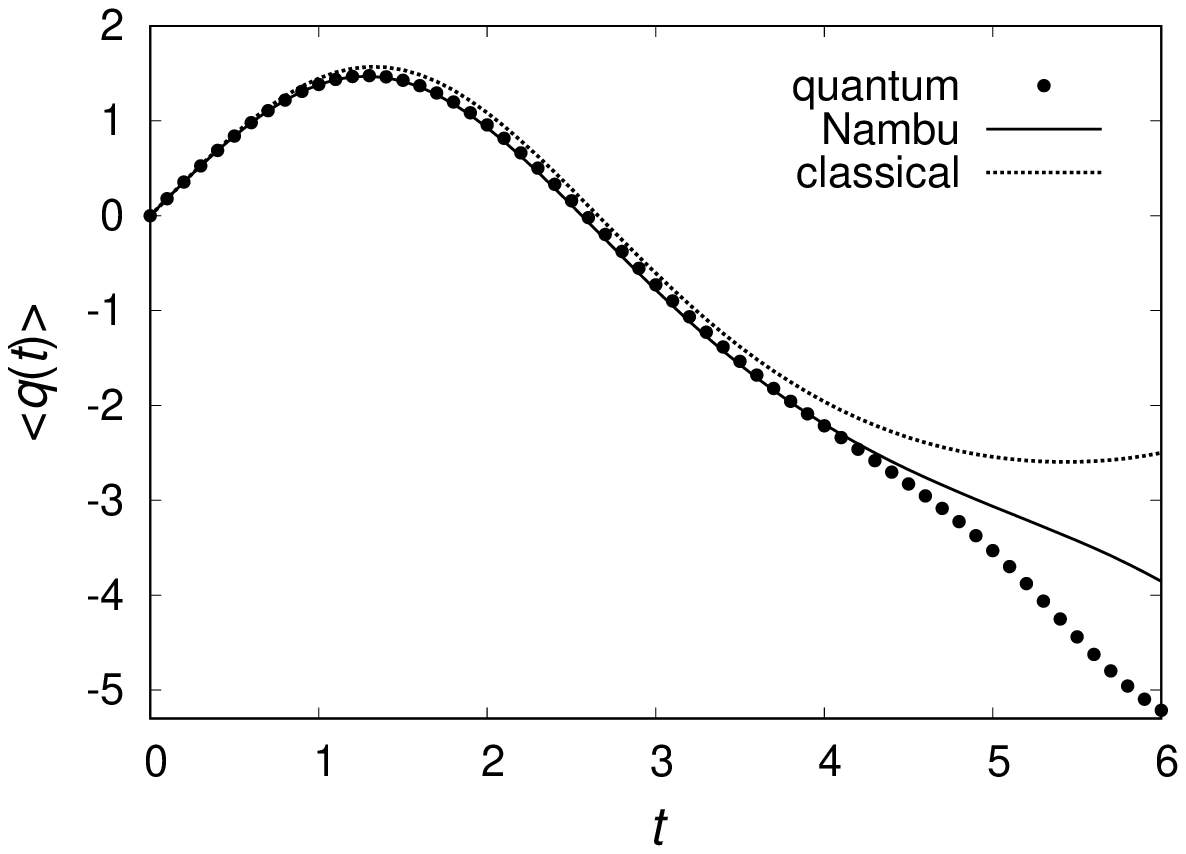}
  \vspace{0mm}
  \hspace{-15mm}
  \subcaption{Calculated trajectories.}
  \label{fig1b}
  \end{center}
 \end{minipage}
 \caption{(a) Classical potential $V(q)$ (thick solid line)
  and initial Gaussian wave packet $|\psi(q,0)|^{2}=|\psi_{\rm FG}(q,0)|^{2}$
  with $(q_{c}(0),p_{c}(0))=(0,1.8)$ (thin solid line).
  The effective potential $V_{c}(q_{c})$, Eq. (\ref{epote}), is also plotted as the dashed line.
  (b) Trajectories in the metastable cubic potential system.
  The quantum trajectory $\langle\hat{q}(t)\rangle$, Nambu trajectory $x_{1}(t)$, 
  and classical trajectory $q(t)$ are given by
  the dots, solid line, and dashed line, respectively.}
 \label{fig1ab}
\end{figure}

The first model is a quantum system which exhibits tunneling.
Consider a one-dimensional metastable system whose quantum Hamiltonian is given by\footnote{
This metastable system has also been used in the applications of 
the symplectic semiclassical wave packet dynamics \cite{OhsawaLeok}.
}
\begin{eqnarray}
\hat{H}=\frac{1}{2m}\hat{p}^{2}+\frac{m\omega^{2}}{2}\hat{q}^{2}+\frac{g}{3}\hat{q}^{3}.
\label{cp}
\end{eqnarray}
The corresponding classical Hamiltonian is $H=(1/2m)p^{2}+V(q)$, where $V(q)$ is the classical potential,
$V(q)=(m\omega^{2}/2)q^{2}+(g/3)q^{3}$.
We choose $N=4$ Nambu variables, as in Eq. (\ref{Ex2q}), 
and the Nambu Hamiltonian $F$ is then determined  by the zero-cumulant approximation in Eq. (\ref{zca}),
\begin{eqnarray}
F=\frac{1}{2m}x_{4}+\frac{m\omega^2}{2}x_{3}+\frac{g}{3}(3x_{3}x_{1}-2x_{1}^{3}).
\label{Nr1-F}
\end{eqnarray}
For the frozen Gaussian wave packet dynamics, the Nambu Hamiltonians
$F$ and $(G_{1}, G_{2})$, Eqs. (\ref{Ex2-G1qFG}) and (\ref{Ex2-G2qFG}),
are conserved in the time evolution according to the semiclassical equations of Eq. (\ref{Ex2-Hq-eqs}),
\begin{eqnarray}
&&~\frac{d}{dt}\langle\hat{q}\rangle=\frac{1}{m}\langle\hat{p}\rangle,~~~~~~~~~~~~~~
\frac{d}{dt}\langle\hat{p}\rangle=-m\omega^{2}\langle\hat{q}\rangle-g\langle\hat{q}^{2}\rangle,\nonumber\\
&&\frac{d}{dt}\langle\hat{q}^{2}\rangle\simeq\frac{2}{m}\langle\hat{q}\rangle\langle\hat{p}\rangle,~~~~~~~~~
\frac{d}{dt}\langle\hat{p}^{2}\rangle\simeq
-2\left(m\omega^{2}\langle\hat{q}\rangle+g\langle\hat{q}^{2}\rangle\right)\langle\hat{p}\rangle.
\label{Nr1-Hq-eqs}
\end{eqnarray}
which are equivalent to the $N=4$ Nambu equations of Eq. (\ref{Ex2-HN-eqs}).
The initial conditions for the Nambu mechanics are given as follows:
\begin{eqnarray}
&&
x_{1}(0)=q_{c}(0),~~~~~~~~~~~~~~~~~~~
x_{2}(0)=p_{c}(0),\nonumber\\
&&
x_{3}(0)=q_{c}^{2}(0)+\frac{\hbar}{2m\omega},~~~~~~~~
x_{4}(0)=p_{c}^{2}(0)+\frac{m\hbar\omega}{2}.
\label{Nr1-ic}
\end{eqnarray}
The classical potential $V(q)$ is plotted in Fig. \ref{fig1a},
where we set the parameters $\omega=1$ and $g=0.3$ with the units $\hbar=m=1$.
These parameters are the same as in Ref. \cite{PrezhdoPereverzev}. 
The initial wave function given by $|\psi(q,0)|^{2}=|\psi_{\rm FG}(q,0)|^{2}$
is also shown in Fig. \ref{fig1a}, where we choose 
the initial conditions as $(q_{c}(0),p_{c}(0))=(0,1.8)$. 
The initial wave packet is located at the local minimum of the classical potential $V(q)$
and moves to the right.
\par
\vspace{0mm}
The calculated trajectories of the quantum, Nambu, and classical mechanics are shown in Fig. \ref{fig1b}.
The quantum mechanical expectation value $\langle\hat{q}(t)\rangle$ 
moves to the right, bounces off the wall, and moves to the left through the potential barrier,
the top of which is located at $q=-3.3$. 
This is an instance of quantum mechanical tunneling because the classical variable $q(t)$ 
fails to go through the potential barrier and oscillates around the local minimum of $V(q)$.
On the other hand, the Nambu variable $x_{1}(t)$ can reproduce the quantum mechanical tunneling, 
although it deviates from the quantum result as time increases.
This semiclassical behavior of the Nambu variable can be understood as follows.
The $N=4$ Nambu mechanics discussed here is equivalent to the variational dynamics of $(q_{c},p_{c})$,
whose time evolution is given by Eq. (\ref{TDVPeqs}).
The effective Hamiltonian is
$H_{c}(q_{c},p_{c})=(1/2m)p_{c}^{2}+V_{c}(q_{c})$, 
where $V_{c}(q_{c})$ is the effective potential shown in Fig. \ref{fig1a},
\begin{eqnarray}
V_{c}(q_{c})=\frac{m\omega^{2}}{2}q_{c}^{2}+\frac{g}{3}q_{c}^{3}
+\frac{\hbar g}{2m\omega}q_{c}+\frac{\hbar\omega}{2}.
\label{epote}
\end{eqnarray}
The last two terms are proportional to the Planck constant $\hbar$ and generated by the quantum correction.
As shown in Fig. \ref{fig1a}, these terms lower the height of the potential barrier,
and there exists a region of initial values $(q_{c}(0),p_{c}(0))$ 
where the Nambu mechanics can tunnel but the classical mechanics cannot. 
The initial conditions adopted here, $(q_{c}(0),p_{c}(0))=(0,1.8)$, are in such a region.

\subsection{Simplified Henon--Heiles model}
\label{Simplified Henon--Heiles model}

The second model is a quantum system which exhibits nonlinear energy exchange dynamics between coupled oscillators.
Consider a one-dimensional quantum system of two oscillators whose Hamiltonian is given by
\begin{eqnarray}
\hat{H}=
\frac{1}{2m_{1}}\hat{p}_{1}^{2}+\frac{1}{2m_{2}}\hat{p}_{2}^{2}
+\frac{m_{1}\omega_{1}^{2}}{2}\hat{q}_{1}^{2}
+\frac{m_{2}\omega_{2}^{2}}{2}\hat{q}_{2}^{2}
+\lambda~\hat{q}_{1}\hat{q}_{2}^{2}.
\label{sHH}
\end{eqnarray}
This is a simplified version of the quantum Henon--Heiles model \cite{PrezhdoPereverzev,HellerStechelDavis}.
We choose $N=4$ Nambu variables as follows:
\begin{eqnarray}
x_1^{(\alpha)}=\langle\hat{q}_{\alpha}\rangle,~~~ x_2^{(\alpha)}=\langle\hat{p}_{\alpha}\rangle,
~~~ x_3^{(\alpha)}=\langle\hat{q}_{\alpha}^{2}\rangle,
~~~ x_4^{(\alpha)}=\langle\hat{p}_{\alpha}^{2}\rangle
~~~ (\alpha=1,2).
\label{Nr2}
\end{eqnarray}
The Nambu Hamiltonian $F$ is determined  by the zero-cumulant approximation of Eq. (\ref{zca}),
\begin{eqnarray}
F=\frac{1}{2m_{1}}x_4^{(1)}+\frac{1}{2m_{2}}x_4^{(2)}
+\frac{m_{1}\omega_{1}^{2}}{2}x_3^{(1)}+\frac{m_{2}\omega_{2}^{2}}{2}x_3^{(2)}
+\lambda~x_1^{(1)}x_3^{(2)}.
\label{Nr2-F}
\end{eqnarray}
In the case of the frozen Gaussian wave packet dynamics, the wave function is approximated by 
the product of two Gaussian wave packets with fixed widths $\sigma_{1}$ and $\sigma_{2}$,
\begin{eqnarray}
\psi_{\rm FG}(q^{(1)},q^{(2)},t)=\prod_{\alpha=1}^{2}
C_{\alpha}~
{\rm exp}\left[
-\frac{1}{4\sigma_{\alpha}^{2}}\left(q^{(\alpha)}-q^{(\alpha)}_{c}(t)\right)^{2}
+\frac{i}{\hbar}p^{(\alpha)}_{c}(t)\left(q^{(\alpha)}-q^{(\alpha)}_{c}(t)\right)\right],
\label{Ex2-FGWP2}
\end{eqnarray}
where $C_{\alpha}=\left(2\pi\sigma_{\alpha}^{2}\right)^{-1/4}$.
Then, $G_{1}^{(\alpha)}$ and $G_{2}^{(\alpha)}$ can be written as
\begin{eqnarray}
&&G_{1}^{(\alpha)}=x_3^{(\alpha)}-(x_1^{(\alpha)})^{2}=\sigma^{2}_{\alpha},
\label{Nr2-G1}\\
&&G_{2}^{(\alpha)}=x_4^{(\alpha)}-(x_2^{(\alpha)})^{2}=\frac{\hbar^{2}}{4\sigma^{2}_{\alpha}}
\label{Nr2-G2}
\end{eqnarray}
$(\alpha=1,2)$.
The Nambu Hamiltonians $F$, $G_{1}=G_{1}^{(1)}+G_{1}^{(2)}$, and $G_{2}=G_{2}^{(1)}+G_{2}^{(2)}$
are all conserved in the following approximated dynamics:
\begin{eqnarray}
&&\frac{d}{dt}\langle\hat{q}_{1}\rangle=\frac{1}{m_{1}}\langle\hat{p}_{1}\rangle,~~~~~~~~~~~~
\frac{d}{dt}\langle\hat{p}_{1}\rangle=
-m_{1}\omega_{1}^{2}\langle\hat{q}_{1}\rangle-\lambda \langle\hat{q}_{2}^{2}\rangle,\nonumber\\
&&\frac{d}{dt}\langle\hat{q}_{1}^{2}\rangle\simeq
\frac{2}{m_{1}}\langle\hat{q}_{1}\rangle\langle\hat{p}_{1}\rangle,~~~~~~~
\frac{d}{dt}\langle\hat{p}_{1}^{2}\rangle\simeq
-2\left(m_{1}\omega_{1}^{2}\langle\hat{q}_{1}\rangle+\lambda \langle\hat{q}_{2}^{2}\rangle\right)
\langle\hat{p}_{1}\rangle,\nonumber\\
&&\frac{d}{dt}\langle\hat{q}_{2}\rangle=\frac{1}{m_{2}}\langle\hat{p}_{2}\rangle,~~~~~~~~~~~~
\frac{d}{dt}\langle\hat{p}_{2}\rangle\simeq
-m_{2}\omega_{2}^{2}\langle\hat{q}_{2}\rangle-2\lambda\langle\hat{q}_{1}\rangle\langle\hat{q}_{2}\rangle,\nonumber\\
&&\frac{d}{dt}\langle\hat{q}_{2}^{2}\rangle\simeq
\frac{2}{m_{2}}\langle\hat{q}_{2}\rangle\langle\hat{p}_{2}\rangle,~~~~~~~
\frac{d}{dt}\langle\hat{p}_{2}^{2}\rangle\simeq
-2\left(m_{2}\omega_{2}^{2}\langle\hat{q}_{2}\rangle+2\lambda\langle\hat{q}_{1}\rangle\langle\hat{q}_{2}\rangle\right)
\langle\hat{p}_{2}\rangle,
\label{Nr2-Hq-eqs}
\end{eqnarray}
which can be written in the $N=4$ Nambu form,
\begin{eqnarray}
\frac{d}{dt}x_i^{(\alpha)} = 
\{x_i^{(\alpha)}, F, G_1, G_{2}\}_{\mbox{\tiny{NB}}} =
\sum_{{\beta}=1}^{2}~
\frac{\partial (x_i^{(\alpha)}, F, G_{1}, G_{2})}
{\partial (x_1^{(\beta)}, x_2^{(\beta)}, x_3^{(\beta)}, x_4^{(\beta)})}.
\label{Nr2-HN-eqn1}
\end{eqnarray}
These equations are equivalent to semiclassical equations of motion
derived from the time-dependent variational principle of Eq. (\ref{TDVP})
using the frozen Gaussian wave packets $\psi_{\rm FG}(q^{(1)},q^{(2)},t)$ in Eq. (\ref{Ex2-FGWP}) 
as a trial wave function.

The initial wave function is given by $\psi_{\rm FG}(q^{(1)},q^{(2)},0)$, 
and corresponding initial conditions for the Nambu mechanics are given 
using the parameterization $\sigma_{\alpha}^{2}=\hbar/(2m_{\alpha}\omega_{\alpha})$,
\begin{eqnarray}
&&
x_{1}^{(\alpha)}(0)=q_{c}^{(\alpha)}(0),~~~~~~~~~~~~~~~~~~~~~~~~~
x_{2}^{(\alpha)}(0)=p_{c}^{(\alpha)}(0),\nonumber\\
&&
x_{3}^{(\alpha)}(0)=(q_{c}^{(\alpha)}(0))^{2}+\frac{\hbar}{2m_{\alpha}\omega_{\alpha}},~~~~~~~~
x_{4}^{(\alpha)}(0)=(p_{c}^{(\alpha)}(0))^{2}+\frac{m_{\alpha}\hbar\omega_{\alpha}}{2}
\label{Nr2-ic}
\end{eqnarray}
$(\alpha=1,2)$.
With the units $\hbar=m_{1}=m_{2}=1$, 
we set the parameters to the same values as in Refs. \cite{PrezhdoPereverzev,HellerStechelDavis}: 
$\omega_{1}=1$, $\omega_{2}=1.1$, and $\lambda=-0.11$.
We choose the initial conditions $(q_{c}^{(1)}(0),p_{c}^{(1)}(0))=(0,0)$ and $(q_{c}^{(2)}(0),p_{c}^{(2)}(0))=(1,1)$,
and then the two wave packets move in harmonic potentials while weakly interacting with each other.

\begin{figure}[htbp]
\hspace{-10mm}
 \begin{minipage}{0.5\hsize}
  \begin{center}
  \includegraphics[width=70mm]{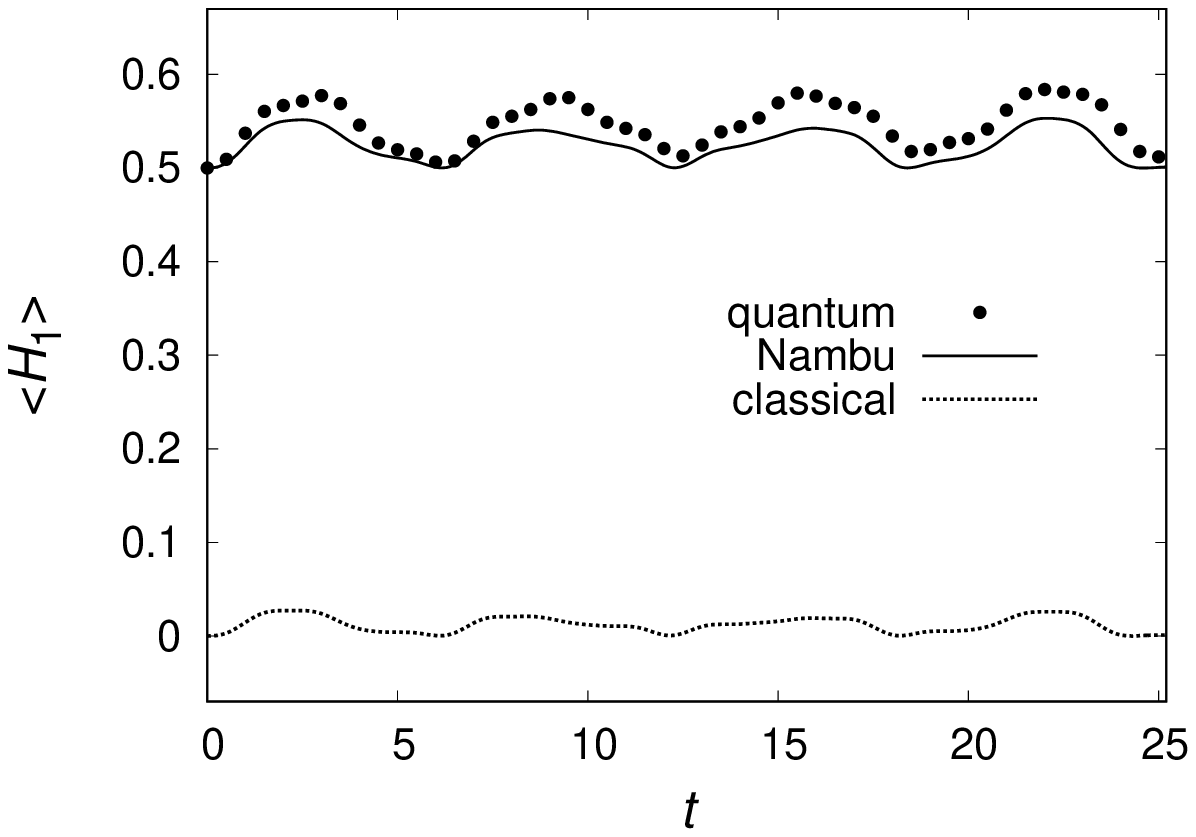}
  \vspace{0mm}
  \hspace{-15mm}
  \subcaption{Harmonic mode energy $\langle\hat{H}_{1}\rangle$.}
  \label{fig2a}
  \end{center}
 \end{minipage}
 \begin{minipage}{0.5\hsize}
  \begin{center}
  \includegraphics[width=70mm]{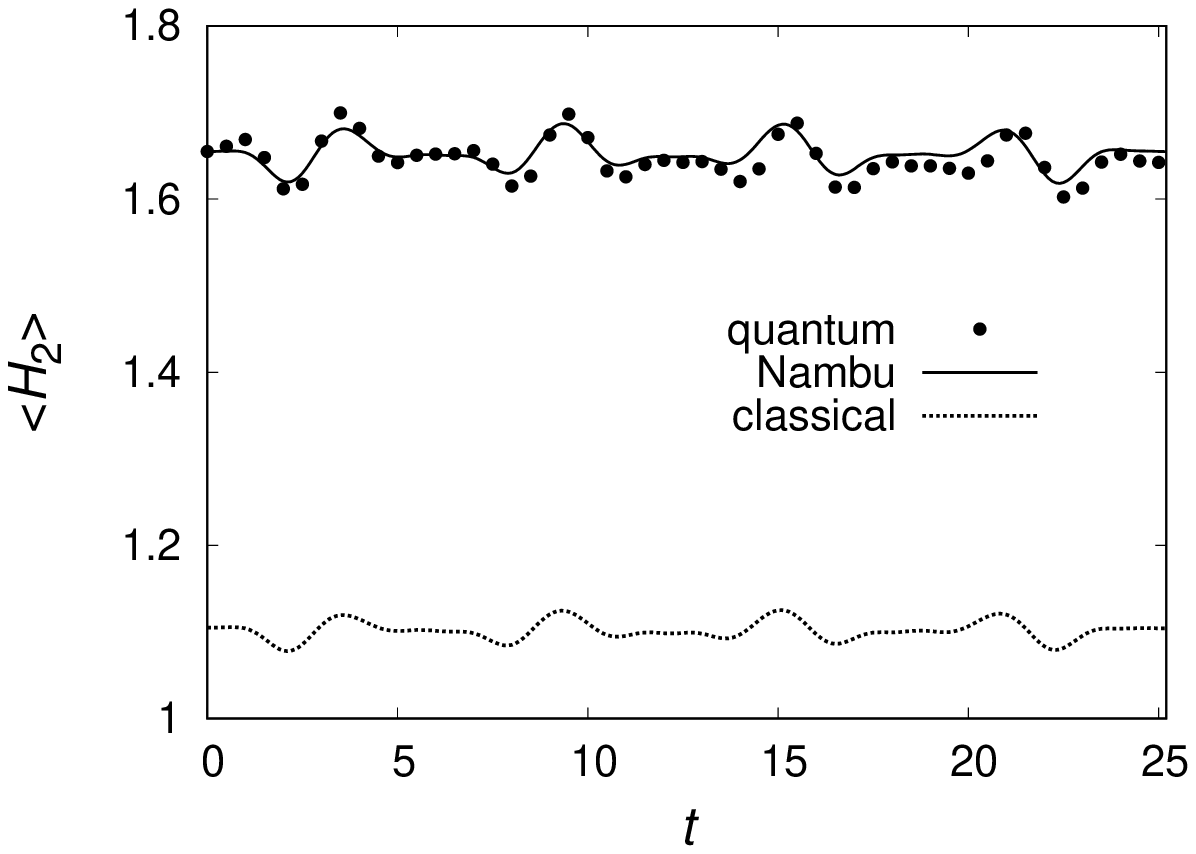}
  \vspace{0mm}
  \hspace{-15mm}
  \subcaption{Harmonic mode energy $\langle\hat{H}_{2}\rangle$.}
  \label{fig2b}
  \end{center}
 \end{minipage}
 \caption{Harmonic mode energies, Eq. (\ref{hme}), of the simplified Henon--Heiles model with the initial conditions 
 $(q_{c}^{(1)}(0),p_{c}^{(1)}(0))=(0,0)$ and $(q_{c}^{(2)}(0),p_{c}^{(2)}(0))=(1,1)$.
  The results for quantum, Nambu, and classical mechanics
  are given by the dots, solid line, and dashed line, respectively.}
 \label{fig2ab}
\end{figure}

Here, we are interested in the dynamics of the harmonic mode energies defined as
\begin{eqnarray}
\langle\hat{H}_{\alpha}\rangle=
\frac{1}{2m_{\alpha}}\langle\hat{p}_{\alpha}^{2}\rangle+
\frac{m_{\alpha}\omega_{\alpha}^{2}}{2}\langle\hat{q}_{\alpha}^{2}\rangle
~~~(\alpha=1,2).
\label{hme}
\end{eqnarray}
The calculated harmonic mode energies for quantum, Nambu, and classical mechanics 
are shown in Figs. \ref{fig2a} and \ref{fig2b}.
The Nambu results are consistent with the quantum mechanical results at $t=0$,
that is, 
the hidden Nambu mechanics can accurately capture the effect of the zero-point energy. 
Although the Nambu results slightly deviate from the quantum results as time increases,
they can reproduce the energy exchange between two oscillators.
On the other hand, due to the lack of the zero-point energy effect,
the classical results are very different from the quantum mechanical results.

\par
\vspace{0mm}

It should be noted here that the Nambu dynamics for two degrees of freedom, Eq. (\ref{Nr2-HN-eqn1}), 
is quite pathological, because, as mentioned in Sect. 4.1, 
the Nambu bracket used in Eq. (\ref{Nr2-HN-eqn1}) 
fails to satisfy the fundamental identity in Eq. (\ref{FI}) for $N=4$.
For example, for $(A_{1},A_{2},A_{3},A_{4})=(x_{1}^{(2)},x_{2}^{(2)},x_{2}^{(1)},x_{4}^{(2)})$
and $(B_{1},B_{2},B_{3})=(F,G_{1},G_{2})$,
the left-hand side of Eq. (\ref{FI}) is zero, whereas the right-hand side is $-\lambda$.
That is, the interaction between two oscillators violates the fundamental identity
and the canonical structure is broken in the hidden Nambu mechanics. 
However, by explicitly solving the constraints in Eqs. (\ref{Nr2-G1}) and (\ref{Nr2-G2}), 
this Nambu mechanics can be reduced to the effective Hamiltonian dynamics
where the time evolution of the canonical doublets can be properly defined.
Therefore the dynamics of two oscillators considered here  is anomalous as the Nambu mechanics,
but not anomalous as the Hamiltonian dynamics. 

\section{Conclusions and future work}
\label{Conclusions and future work}

We have shown that the Nambu mechanical structure is hidden 
not only in classical Hamiltonian dynamics
but also in some quantum or semiclassical dynamics.
We focused on the dynamics defined in an extended phase space 
spanned by $N(\ge 3)$ quantum mechanical expectation values.\footnote{
Our formalism could also be applied to statistical-mechanical expectation values.}
The dynamics of variables such as 
$(\langle\hat{q}\rangle,~\langle\hat{p}\rangle,~\langle\hat{q}^{2}\rangle,~\langle\hat{p}^{2}\rangle)$
cannot be described by the Hamilton equations of motion; however,
if the system has a sufficient number of conserved quantities, $(F,G_{1},G_{2})$, 
their dynamics could be described by the $N=4$ Nambu equations.
We gave some quantum/semiclassical examples of hidden Nambu mechanics, 
including a many-degrees-of-freedom system.
It would be interesting to investigate other examples.

In many-degrees-of-freedom systems, however, the hidden Nambu mechanics become anomalous,
because interactions between multiple degrees of freedom 
violate the fundamental identity of Eq. (\ref{FI}) \cite{Takhtajan,HoMatsuo}.
Since the fundamental identity would play a similar role to the Jacobi identity in the Hamiltonian dynamics,
its violation implies that 
it would be difficult to formulate the Nambu statistical mechanics or quantize the Nambu mechanics.
On the other hand, in Hamiltonian dynamics there also exists anomalous dynamics
known as nonholonomic dynamics \cite{BlochMarsdenZenkov},
where the Jacobi identity is violated and the Hamiltonian structure is broken 
\cite{SatoYoshida}. 
Recently, a procedure has been proposed to recover the Hamiltonian structure 
and formulate a statistical theory of the nonholonomic dynamics \cite{Sato}.
This work might provide guidance for formulating a statistical theory of Nambu mechanics,
and our formalism presented in this article might provide example systems
suitable for Nambu statistical mechanics to be tested.

\section*{Acknowledgments}
We would like to thank Yoshiharu Kawamura for helpful comments,
and William H. Miller for suggesting the relationship between the hidden Nambu mechanics 
and the quantized Hamiltonian dynamics.
This work was supported in part by a scientific grant 
from the Ministry of Education, Culture,
Sports, Science and Technology under Grant No.~26610134.



\begin{thebibliography}{99}

\bibitem{Nambu} Y.~Nambu, 
Phys. Rev. D~{\bf 7}, 2405 (1973).

\bibitem{Takhtajan} L.~Takhtajan, 
Commun. Math. Phys. {\bf 160}, 295 (1994).

\bibitem{DitoFlatoSternheimerTakhtajan} G.~Dito, M.~Flato, D.~Sternheimer, and L.~Takhtajan, 
Commun. Math. Phys. {\bf 183}, 1 (1997).

\bibitem{AwataLiMinicYoneya} H.~Awata, M.~Li, D.~Minic, and T.~Yoneya, 
J. High Energy Phys. {\bf 0102}, 013 (2001).

\bibitem{CurtrightZachos} T.~Curtright and C.~Zachos,
Phys. Rev. D~{\bf 68}, 085001 (2003).

\bibitem{AxenidesFloratosNicolis} M.~Axenides, E.~G.~Floratos, and S.~Nicolis,
J. Phys. A: Math. Theor. {\bf 42}, 275201 (2009).   

\bibitem{deAzcarragaIzquierdo} J.~A.~de Azcarraga and J.~M.~Izquierdo,
J. Phys. A: Math. Theor. {\bf 43}, 293001 (2010).

\bibitem{MongkolsakulvongChaikhanFrank} S.~Mongkolsakulvong, P.~Chaikhan, and T.~D.~Frank,
Eur. Phys. J. B~{\bf 85}, 90 (2012).

\bibitem{BlenderLucarini} R.~Blender and V.~Lucarini
Physica D~{\bf 243}, 86 (2013).

\bibitem{SaitouBambaSugamoto} M.~Saitou, K.~Bamba, and A.~Sugamoto,
Prog. Theor. Exp. Phys. {\bf 2014}, 103B03 (2014).

\bibitem{HoMatsuo} P.-M.~Ho and Y.~Matsuo,
Prog. Theor. Exp. Phys. {\bf 2016}, 06A104 (2016).

\bibitem{Yoneya} T.~Yoneya,
Prog. Theor. Exp. Phys. {\bf 2017}, 023A01 (2017).

\bibitem{HorikoshiKawamura} A.~Horikoshi and Y.~Kawamura,
Prog. Theor. Exp. Phys. {\bf 2013}, 073A01 (2013).

\bibitem{Heller} E.~J.~Heller, 
J. Chem. Phys. {\bf 75}, 2923 (1981).

\bibitem{SatoYoshida} N.~Sato and Z.~Yoshida, 
Phys. Rev. E~{\bf 97}, 022145 (2018).

\bibitem{PrezhdoPereverzev} O.~V.~Prezhdo and Y.~V.~Pereverzev,
J. Chem. Phys. {\bf 113}, 6557 (2000).

\bibitem{Prezhdo} O.~V.~Prezhdo,
J. Chem. Phys. {\bf 117}, 2995 (2002).

\bibitem{FeldmeierSchnack} H.~Feldmeier and J.~Schnack,
Rev. Mod. Phys. {\bf 72}, 655 (2000).

\bibitem{Modin} K.~Modin, 
J. Gen. Lie Theory Appl. {\bf 3}, 39 (2009).

\bibitem{WatanabeTsukada} N.~Watanabe and M.~Tsukada, 
Phys. Rev. E~{\bf 62}, 2914 (2000).

\bibitem{OhsawaLeok} T.~Ohsawa and M.~Leok,
J. Phys. A: Math. Theor. {\bf 46}, 405201 (2013).

\bibitem{HellerStechelDavis} E.~J.~Heller, E.~B.~Stechel, and M.~J.~Davis, 
J. Chem. Phys. {\bf 73}, 4720 (1980).

\bibitem{BlochMarsdenZenkov} A.~M.~Bloch, J.~E.~Marsden, and D.~V.~Zenkov, 
Not. AMS {\bf 52}, 320 (2005).

\bibitem{Sato} N.~Sato, 
arXiv:1806.05003 [math-ph].

\end{thebibliography}
\end{document}